\documentclass[sigconf]{acmart}

\AtBeginDocument{%
  }

\copyrightyear{2023}
\acmYear{2023}
\setcopyright{acmlicensed}
\acmConference[RAID '23]{The 26th International Symposium on Research in Attacks, Intrusions and Defenses}{October 16--18, 2023}{Hong Kong, China}
\acmBooktitle{The 26th International Symposium on Research in Attacks, Intrusions and Defenses (RAID '23), October 16--18, 2023, Hong Kong, China}
\acmPrice{15.00}
\acmDOI{10.1145/3607199.3607231}
\acmISBN{979-8-4007-0765-0/23/10}

\newcommand\Tstrut{\rule{0pt}{2.4ex}}
\usepackage{color}

\usepackage{tikz}
\newcommand\copyrightnotice[1]{
	\begin{tikzpicture}[remember picture,overlay]
		\node[anchor=south,yshift=20pt] at (current page.south) {\fbox{\parbox{\dimexpr\textwidth-\fboxsep-\fboxrule\relax}{#1}}};
	\end{tikzpicture}
}

\begin{document}

%%
%% The "title" command has an optional parameter,
%% allowing the author to define a "short title" to be used in page headers.
\title[Leveraging XAI to Analyze the Reasoning and True Performance of DGA Classifiers]{False Sense of Security: Leveraging XAI to Analyze the Reasoning and True Performance of Context-less DGA Classifiers}

%%
%% The "author" command and its associated commands are used to define
%% the authors and their affiliations.
%% Of note is the shared affiliation of the first two authors, and the
%% "authornote" and "authornotemark" commands
%% used to denote shared contribution to the research.
\author{Arthur Drichel}
\email{drichel@itsec.rwth-aachen.de}
\affiliation{%
			\institution{RWTH Aachen University}
			\city{}
			\country{}
			%	\city{Aachen}
			%	\country{Germany}
		}
\author{Ulrike Meyer}
\email{meyer@itsec.rwth-aachen.de}
\affiliation{%
			\institution{RWTH Aachen University}
			\city{}
			\country{}
		}

%%
%% By default, the full list of authors will be used in the page
%% headers. Often, this list is too long, and will overlap
%% other information printed in the page headers. This command allows
%% the author to define a more concise list
%% of authors' names for this purpose.
%\renewcommand{\shortauthors}{Drichel and Meyer}
%\renewcommand{\shortauthors}{Drichel et al.}

%%
%% The abstract is a short summary of the work to be presented in the
%% article.
% !TEX root = ../paper.tex
\begin{abstract}
The problem of revealing botnet activity through Domain Generation Algorithm (DGA) detection seems to be solved, considering that available deep learning classifiers achieve accuracies of over 99.9\%.
However, these classifiers provide a false sense of security as they are heavily biased and allow for trivial detection bypass.
In this work, we leverage explainable artificial intelligence (XAI) methods to analyze the reasoning of deep learning classifiers and to systematically reveal such biases.
We show that eliminating these biases from DGA classifiers considerably deteriorates their performance.
Nevertheless we are able to design a context-aware detection system that is free of the identified biases and maintains the detection rate of state-of-the art deep learning classifiers. 
In this context, we propose a visual analysis system that helps to better understand a classifier's reasoning, thereby increasing trust in and transparency of detection methods and facilitating decision-making.
\end{abstract}

%%
%% The code below is generated by the tool at http://dl.acm.org/ccs.cfm.
%% Please copy and paste the code instead of the example below.
%%


%%
%% Keywords. The author(s) should pick words that accurately describe
%% the work being presented. Separate the keywords with commas.
%% A "teaser" image appears between the author and affiliation
%% information and the body of the document, and typically spans the
%% page.
%\begin{teaserfigure}
%  \includegraphics[width=\textwidth]{sampleteaser}
%  \caption{Seattle Mariners at Spring Training, 2010.}
%  \Description{Enjoying the baseball game from the third-base
%  seats. Ichiro Suzuki preparing to bat.}
%  \label{fig:teaser}
%\end{teaserfigure}

%\received{20 February 2007}
%\received[revised]{12 March 2009}
%\received[accepted]{5 June 2009}

%%
%% This command processes the author and affiliation and title
%% information and builds the first part of the formatted document.
\maketitle

\copyrightnotice{\copyright\space Copyright held by the owner/author(s) 2023. This is the author's version of the work. It is posted here for your personal use. Not for redistribution. The definitive version was published in The 26th International Symposium on Research in Attacks, Intrusions and Defenses (RAID ’23), https://doi.org/10.1145/3607199.3607231} 

% !TEX root = ../paper.tex
\section{Introduction}
\label{sec:introduction}

In recent years, deep learning has been increasingly used as a building block for security systems incorporating classifiers that achieve high accuracies in various classification tasks.
The advantage of deep learning classifiers is that they often outperform classical machine learning approaches, can be trained in an end-to-end fashion, and automatically learn to extract relevant features for classification.
Therefore, less effort is often expended in creating such classifiers, since they seem to achieve high accuracies out-of-the-box and do not require the integration of domain knowledge as would be required to create feature-based or rule-based classifiers.

This black-box nature of deep learning classifiers is particularly dangerous in the security domain, as the classifiers operate in an adversarial environment where an attacker actively aims to avoid detection.
Since it is unclear what a classifier has learned, not only is its operation opaque, leading to trust issues, but it is also unclear whether the training data might have influenced a classifier in a way that an attacker could easily bypass the classification.
Related work~\cite{arp_dos_2022,cortes_sample_2008, lipton_detecting_2018, pendlebury_tesseract_2019, axelsson_baserate_2000} has identified and summarized common pitfalls when using machine learning in computer security, including pitfalls that make it easier for an attacker to evade detection.
These pitfalls range from sampling bias, where the data used does not adequately represent the true data distribution, over inaccurate ground-truth labels, to incorporating spurious correlations, where artifacts unrelated to the classification problem provide shortcuts for distinguishing classes.
To uncover potential classification biases introduced by these pitfalls, related work suggests using explainability techniques for machine learning.
However, it remains unclear which strategy is appropriate to mitigate identified problems.

In this work, we systematically apply explainability techniques to the use-case of Domain Generation Algorithm (DGA) detection to reveal a variety of biases in state-of-the-art deep learning classifiers.
We then evaluate the loss in classification performance induced by the elimination of these biases from the classifiers and propose a classification system that is free of the identified biases.

We focus on DGA detection because for this use-case a plethora of research exists, the state-of-the-art classifiers that achieve accuracies up to 99.9\% are open source, and domains generated by different DGAs are publicly available in bulk through open source intelligence (OSINT) feeds such as DGArchive~\cite{plohmann_comprehensive_2016}. This allows us to replicate the results of related work before performing a critical analysis of automatic feature extraction.

To this end, we first conduct an extensive evaluation of a variety of different explainability techniques including recent developments.
Then, we demonstrate how these methods can be used to debug and improve the understanding of state-of-the-art classifiers.
In this context, we identify features and classification biases and show how this knowledge can be exploited to evade detection with ease.
To address these issues, we propose a classification system free of the identified biases combined with a visualization system that supports analysts in Security Operation Centers (SOCs), increases transparency and confidence in detection methods, and facilitates decision-making.

Finally, as a secondary contribution, we use the knowledge gained from our study to improve the state-of-the-art deep learning as well as feature-based approaches for DGA multiclass classification in terms of classification performance and efficiency.

Overall, we thus provide a systematic approach to expose biases and analyze the reasoning of deep learning classifiers for DGA detection.
While some of these biases may seem obvious and easily avoidable, they are present even in DGA detection approaches proposed at leading security conferences (e.g.,~\cite{schuppen_fanci_2018}).
Moreover, these biases are rooted on subtle flaws that are rife in security research and affect many other use-cases as well~\cite{arp_dos_2022}.
Thus, with this work we aim to raise awareness of potential pitfalls in state-of-the-art classifiers that allow bypassing detection, and provide helpful guidance in conducting a similar analysis also for different use-cases.
While features and biases are highly domain specific, the generation of explanations is completely independent of the underlying classification task.
Hence, the fundamental idea of leveraging XAI to improve machine learning classifiers is applicable to a variety of different use-cases (e.g., phishing detection, malware detection, vulnerability discovery, or general network intrusion detection).

% !TEX root = ../paper.tex
\section{Preliminaries}
\label{sec:dga_detection}
The self-learned features of a deep learning classifier and thus potential biases in its classification decision are mostly use-case dependent.
It is thus fundamental to understand the specifics of the classification task at hand, including the data used by state-of-the-art classifiers and the data preprocessing applied.

\subsection{Domain Generation Algorithm Detection}

Domain Generation Algorithms (DGAs) are used by malware infected devices to contact the botnet master's command and control (C2) server for updates or instructions (e.g., the target IP for a distributed denial-of-service (DDoS) attack).
DGAs are pseudo-random algorithms which generate a large amount of domain names that the bots query one by one.
The advantage of this approach over using fixed IP addresses or fixed domain names is that it creates an asymmetric situation where the botnet master only needs to register one domain, but the defenders have to block all generated domains.
The botnet master knows the seed and the generation scheme and can thus register a DGA-generated domain in advance.
When the bots query this domain, they get the valid C2 server's address, while all other queries result in non-existent domain (NXD) responses.

\subsection{State-of-the-Art Classifiers}
To combat DGAs, binary detection approaches have been proposed in the past, capable of distinguishing benign domains from DGA-generated domains with high probability and low false-positive rates (e.g.,~\cite{drichel_analyzing_2020, schuppen_fanci_2018, woodbridge_predicting_2016,yu_character_2018}).
Going a step further, multiclass classifiers have been proposed that can not only separate benign domains from DGA-generated domains, but are also able to associate malicious domains with the DGA that generated them, allowing for the identification and targeted remediation of malware families (e.g.,~\cite{drichel_analyzing_2020,drichel_first_2021,tran_lstm_2018,woodbridge_predicting_2016}).

In general these approaches can be divided into two groups: context-less (e.g.,~\cite{drichel_analyzing_2020,saxe_expose_2017,schuppen_fanci_2018,woodbridge_predicting_2016,yu_character_2018,tran_lstm_2018}) and context-aware (e.g.,~\cite{antonakakis_throwaway_2012,bilge_exposure_2014,shi_malicious_2018,grill_detecting_2015,schiavoni_phoenix_2014,yadav_winning_2012}) approaches.
Context-less approaches work exclusively with information that can be extracted from a single domain name, while context-aware approaches use additional information, such as statistical data from the monitored network, to further improve detection performance.
Previous studies (e.g.,~\cite{drichel_analyzing_2020,schuppen_fanci_2018,woodbridge_predicting_2016,yu_character_2018}) have shown that context-less approaches achieve similar or even higher performance while requiring less resources and being less intrusive than context aware approaches. 

Furthermore, the machine learning classifiers can additionally be divided into feature-based classifiers such as support vector machines (SVMs) or random forests (RFs) (e.g.,~\cite{schuppen_fanci_2018,drichel_first_2021,bilge_exposure_2014}), and feature-less (deep learning-based) classifiers such as recurrent (RNNs), convolutional (CNNs), or residual neural networks (ResNets) (e.g.,~\cite{drichel_analyzing_2020,woodbridge_predicting_2016,yu_character_2018,saxe_expose_2017}).
Previous studies (e.g.,~\cite{drichel_analyzing_2020,woodbridge_predicting_2016,peck_charbot_2019,spooren_detection_2019,sivaguru_evaluation_2018}) have shown that feature-less approaches achieve superior classification performance.

The currently best deep learning-based classifier for binary and multiclass classification is ResNet~\cite{drichel_analyzing_2020}.
Hence, we analyze the reasoning of this particular classifier in detail.
In addition, we use the insights gained from our analysis to identify missing features in EXPLAIN~\cite{drichel_first_2021}, currently the most powerful feature-based multiclass classifier, and seek to bring its classification performance up to the state-of-the-art level.

In the following, we briefly introduce both classifier types.
Detailed information on the implementations of each classifier can be found in~\cite{drichel_analyzing_2020, drichel_first_2021}.

\subsubsection{ResNet}
Drichel et al.~\cite{drichel_analyzing_2020} proposed ResNet-based models for DGA binary and multiclass classification.
The classifiers are constructed from residual blocks containing skip connections between convolutional layers to counteract the vanishing gradient problem.
\textit{B-ResNet}, the proposed binary classifier, uses only one residual block with 128 filters per convolutional layer while \textit{M-ResNet}, the multiclass classifier, is more complex and composed of eleven residual blocks with 256 filters.

\subsubsection{EXPLAIN}
The authors of EXPLAIN~\cite{drichel_first_2021} proposed several variants of their feature-based and context-less DGA multiclass classifier.
The best performing model is a one-vs.-rest variant of a RF that extracts 76 features for each domain name to be classified, which can be categorized into 51 linguistic, 19 statistical and 6 structural features.

\subsection{Data}
\label{sec:data}
To train machine learning classifiers for DGA classification, domain names labeled with the DGA that generated them are widely available in OSINT feeds such as DGArchive~\cite{plohmann_comprehensive_2016}.
Benign training data can either be obtained by monitoring real networks or generated artificially based on public top sites rankings such as Tranco~\cite{lepochat_tranco_2019}.
The problem with artificial data is that it may not accurately reflect real network traffic and thus may introduce bias and lead to misleading results.
Further, the domain names included in public top sites rankings are on the resolving side of the DNS traffic because they are registered.
Since most DGA-generated domains are not registered, additional bias may be introduced when they are paired with registered benign domain names for training.
Due to these reasons, several approaches (e.g.,~\cite{schuppen_fanci_2018,drichel_first_2021,drichel_analyzing_2020,drichel_making_2020,antonakakis_throwaway_2012,yadav_winning_2012,tong_far_2020}) focus on the classification of non-resolving DNS traffic (NX-traffic).
Moreover, the focus on NX-traffic offers a number of other advantages:
First, NX-traffic is easier to monitor because its volume is an order of magnitude smaller than the volume of full DNS traffic.
Monitoring NX-traffic still allows us to detect malware-infected machines before they are instructed to participate in malicious actions, as DGAs can usually be detected in NX-traffic long before they resolve a registered domain for their C2 server.
Second, NXDs are less privacy-sensitive compared to resolving domain names, as they generally do not contain user-generated domains, with the exception of typo domains.
Although, NXDs may still contain sensitive information about an organization as a whole, the classification of NX-traffic seems better suited to a Classification-as-a-Service (CaaS) setting.
Finally, it has been shown that classifiers trained on NX-traffic are more robust against certain adversarial attacks compared to classifiers trained on resolving traffic~\cite{drichel_analyzing_2020}.

In this work, we follow the suggestions of related works and focus on the classification of NX-traffic.
In the following, we briefly describe our data sources.

\subsubsection{DGArchive}
We use the OSINT feed of DGArchive~\cite{plohmann_comprehensive_2016} to obtain DGA-labeled domains.
At the time of writing the feed contains approximately 123 million unique samples generated by 106 different DGAs.

\subsubsection{University Network}
We extract benign-labeled domain names from traffic recordings of the central DNS resolver of the campus network of RWTH Aachen University.
This network includes several academic and administrative networks, dormitory networks, and the network of the affiliated university hospital.
We selected a one-month recording of NXDs from mid-October 2017 until mid-November 2017 containing approximately 35 million unique NXDs for our evaluation.
We deliberately chose an older NX-traffic recording because in our study we also want to evaluate whether a classifier learns time-dependent artifacts of a specific network or whether it generalizes well to new environments and is time-robust.

We filter all NXDs from this data source using DGArchive to remove potentially malicious domains.
Although the data may still contain mislabeled samples, the only way to avoid this problem is to use artificial data which may not accurately reflect real network traffic and thus may introduce additional bias.

\subsubsection{Company Network}
A second source for benign-labeled data are recordings of several central DNS resolvers of Siemens AG.
Data obtained from this source is very diverse as the DNS resolvers cover the regions of Asia, Europe, and the USA.
From the company, we obtain a one-month recording of benign NXDs from April 2019 containing approximately 311 million unfiltered NXDs.
Benign data from this source is only used for the final real-world evaluation study, which is free of experimental biases, to assess whether a classifier contains any biases with respect to the network data on which it was trained and whether a classifier is time-robust.

We again filter all NXDs from this data source using DGArchive to clean the data as much as possible.

\subsubsection{Ethical Considerations}
Our institution does not yet have an ethics review board that could have approved this study.
However, we ensured that we do not record or use any personally identifiable information (PII) or quasi-identifiers.  
When recording traffic from the university and company network, we only observe NX-traffic and store the queried domain names, omitting all other information including IP addresses that could be used as pseudonyms to correlate domain names queried by the same host.
Thereby, we only obtain a list of domain names that occurred within the recording period, with no relation to users within the network.
Additionally, we focus on NX-traffic because NXDs are less privacy-sensitive compared to resolving domain names, as they generally do not contain user-generated domains, with the exception of typo domains.
Although the NXDs may still contain sensitive information about an organization as a whole (e.g., they could indicate possible business relationships between different companies), it is questionable to what extent and with what accuracy such information can be recovered, if at all possible.

\subsection{Preprocessing}
It is important to understand the applied domain name preprocessing as this step can introduce significant classification biases.
The works (e.g.,~\cite{drichel_analyzing_2020, drichel_first_2021, schuppen_fanci_2018, drichel_making_2020}) that operate on single NXDs for classification make the data used unique and filter all benign samples against OSINT feeds to remove potentially contained malicious domains before training and testing a classifier.
Other than that, they do not apply any filtering to the benign-labeled data used, since it is captured from real-world networks.
The argument for this decision is that this feeds the classifier with the queries that occur naturally in a network, and does not bias the classification performance in any direction since no filtering is applied.
While the feature-based classifiers (e.g.,~\cite{schuppen_fanci_2018, drichel_first_2021}) start extracting predefined features from this data, the deep learning-based approaches (e.g.,~\cite{drichel_analyzing_2020, yu_character_2018, woodbridge_predicting_2016, tran_lstm_2018, saxe_expose_2017}) have to convert the domain names into a numerical representation in order to be able to feed them to a neural network.
Most works (e.g.,~\cite{drichel_analyzing_2020, yu_character_2018, woodbridge_predicting_2016, tran_lstm_2018}) follow a similar approach, which mainly differs in the maximum acceptable length of a domain.
First, all characters are converted to lowercase (which is an uncritical operation as the DNS operates case-insensitive) and every character is mapped to a unique integer.
Additionally, the input is padded with zeros from the left side.
The authors of the ResNet classifier~\cite{drichel_analyzing_2020} propose padding to the maximum domain length of 253 characters in order to be able to perform training and classification on every possible NXD while using batch learning.
In this work, we follow these suggestions of related work on preprocessing.

% !TEX root = ../paper.tex
\section{Evaluation Overview}
\label{sec:methodology}
In this section, we describe our evaluation methodology, explain the decisions underlying the dataset generation process, and perform a result reproduction study of the classifiers from related work to verify our evaluation setup.

\subsection{Datasets \& Methodology}
We create two disjoint datasets, one to train and test a set of state-of-the-art models (\textit{DS\textsubscript{mod}}), and one to analyze different explainability methods and investigate biases (\textit{DS\textsubscript{ex}}).

For each DGA in DGArchive, we randomly select 20,000 samples.
If less than 20,000 samples are available per DGA, we select all samples.
Then we split the samples for each DGA equally between the two datasets.
For two DGAs, only five samples are available in the OSINT feed.
We constrain that at least four samples are available for training classifiers within DS\textsubscript{mod}.
Thus, for two DGAs (\textit{Dnsbenchmark} and \textit{Randomloader}), only one sample is contained in DS\textsubscript{ex}.\footnote{We intentionally include underrepresented classes because the inclusion of a few training samples per class allows a classifier to detect various underrepresented DGAs with high probability that would otherwise be missed. At the same time, this does not affect a classifier's ability to recognize well-represented classes~\cite{drichel_making_2020}.}
Thereby, we are able to perform a four-fold cross validation stratified over all included classes using DS\textsubscript{mod}, resulting in four different classifiers being trained and tested.
Finally, we select the same number of benign samples as we selected malicious samples, resulting in balanced datasets.

In binary classification experiments, we use all benign samples and use the same label for all malicious domains, regardless of which DGA generated a domain.
In multiclass classification experiments, we limit the amount of benign samples to 10,000 in order to have a more evenly distributed amount of samples between the various classes.
Here we assign a separate label for each DGA.

In total, DS\textsubscript{mod} and DS\textsubscript{ex} each contain approximately 1.2 million domains derived from 107 different classes.

We train all four classifiers in the four-fold cross validation with DS\textsubscript{mod} using early stopping with a patience of five epochs to avoid overfitting.
These classifiers are then used to analyze different explainability methods and investigate biases using samples from DS\textsubscript{ex}.

This methodology allows us to conduct a study to reproduce the results of related work (using DS\textsubscript{mod}) as it replicates the classification setting used by the state of the art.
In addition, we can evaluate four classifiers and 20 explainability methods on the same unseen data (DS\textsubscript{ex}) and can assess whether the classifiers converge to similar local optima and whether the explainability methods provide stable results between different models.

However, this methodology introduces spatial and temporal experimental biases~\cite{pendlebury_tesseract_2019}. 
Spatial bias arises from using an unrealistic ratio of benign to malicious samples in the test data. For the DGA detection use-case, most queried domains within a network are benign. This significant class imbalance can lead to base-rate fallacy~\cite{axelsson_baserate_2000} where evaluation metrics such as true-positive rate (TPR) and false-positive-rate (FPR) are misleading.

Temporal bias is introduced by temporally inconsistent evaluations which integrate future knowledge about testing samples into the training phase. In the state-of-the-art classification setting, temporal bias is introduced in two ways: First, four-fold cross validation does not ensure that all training samples are strictly temporally precedent to the testing ones. Second, the benign and malicious samples in the datasets are not from the same time window (one-month real-world benign data compared to several years of DGArchive data).

Thus, we conduct an additional evaluation under real-world conditions where we mitigate all experimental biases in Section~\ref{sec:rw_study}.
To this end, we  make use of our second source for real-world data, the company network. 
In this context, we also assess whether classifiers generalize between different networks and are time-robust.

\subsection{State-of-the-Art Results Reproduction}
\label{sec:sota_result_reproduction}
Before conducting the actual explainability study, we reproduce the results of related work to validate our evaluation setup.
We use the same evaluation metrics as in the original papers: accuracy (ACC), true-positive rate (TPR), and false-positive rate (FPR) for the binary experiments, and f1-score, precision, and recall (which is equal to TPR) for the multiclass experiments.
As suggested in~\cite{drichel_analyzing_2020}, we use macro-averaging to calculate the overall evaluation metrics because the available samples vary widely per DGA class.
This way we do not skew the overall score towards well-represented classes.

\begin{table}
	\caption{Outcome of the result reproduction study.}
	\label{tab:results_reproduction}
	\centering
	\resizebox{\linewidth}{!}{
		\begin{tabular}{lcccc}
			\hline \Tstrut
			\textbf{Model} & \textbf{Setting} & \textbf{ACC} & \textbf{TPR} & \textbf{FPR} \\
			\hline \Tstrut
			B-ResNet~\cite{drichel_analyzing_2020} & Binary & 0.99864 & 0.99982 & 0.00255 \\
			\toprule
			\toprule
			\textbf{Model} & \textbf{Setting} & \textbf{F1-Score} & \textbf{Precision} & \textbf{Recall} \\
			\hline \Tstrut
			M-ResNet~\cite{drichel_analyzing_2020} & Multiclass & 0.78682 & 0.80058 & 0.79690 \\
			EXPLAIN~\cite{drichel_first_2021} & Multiclass & 0.76733 & 0.78604 & 0.76685 \\
			\hline \Tstrut
			M-ResNet~\cite{drichel_analyzing_2020} + B-Cos~\cite{bohle_b-cos_2022} & Multiclass & 0.76990 & 0.79555 & 0.77250 \\
		\end{tabular}
	}
\end{table}

We present the averaged results of the four-fold cross validation in Table~\ref{tab:results_reproduction}.
The upper part of the table shows the results of the binary evaluation, the lower part those of the multiclass evaluation.
By comparing these results with the values reported in the original papers, we can confirm that we were able to reproduce the results, as we arrive at very similar values.

The last row of the table shows the results for an adapted model of M-ResNet aimed at making it more explainable.
Recently, Bohle et al.~\cite{bohle_b-cos_2022} proposed a so-called B-Cos transform which, when interchanged with linear transforms of neural networks, increases the networks' explainability by promoting the alignment of weight-input during training.
The alignment pressure on the weights ensures that the model computations align with task-relevant features and therefore become explainable.
Since interchanging the linear transforms of the ResNet model with B-Cos transforms could introduce a trade-off between classification performance and explanatory fidelity, we also evaluate this model using DS\textsubscript{mod} and present the results in the last row of Table~\ref{tab:results_reproduction}.
Indeed, this modification slightly sacrifices model performance in favor of a more explainable model compared to the M-ResNet baseline.

% !TEX root = ../paper.tex
\section{Explainability Methods}
\label{sec:explainability_methods}

As a secondary contribution to the critical analysis of automatic feature extraction for DGA detection, we conduct a comparative evaluation of different explainability methods.
In this section, we briefly introduce explainability techniques for machine learning and present the results of the comparative evaluation. 
The exhaustive evaluation can be found in Appendix~\ref{sec:evaluating_explainability_methods}.

In general, explainability methods can be divided into two categories: white-box approaches, which are model-specific and use knowledge, e.g, about the internal architecture and model weights of a neural network, and black-box approaches that are model-agnostic.
In this work, we focus on white-box approaches as they have been proven to produce better results compared to black-box approaches~\cite{warnecke_evaluating_2020,atanasova_diagnostic_2020}.

The general idea of white-box approaches to deriving local explanations for input samples is to compute the gradients from the output back to the input.
Thereby, for an input sample \mbox{$x = (x_0,...,x_n)$}, a neural network $N$, and a prediction \mbox{$f_N(x) = y$}, a relevance vector \mbox{$r = (r_0,...,r_n)$} is derived which describes the relevance of each dimension of $x$ for the predicted label $y$.
Thus, in terms of context-less DGA classification, an explainability method determines the relevance of each character in the context of its position for the assignment of an individual domain name to a particular class.

When evaluating the explainability methods, we focus on the explanations generated for the predictions of a multiclass classifier because, unlike a binary classifier, it has a variety of other  prediction possibilities in addition to distinguishing between benign and malicious.

In this work, we make use of the iNNvestigate library~\cite{alber_innvestigate_2019} which implements many explainability methods and provides a common interface to evaluate 19 white-box approaches including Layer-wise Relevance Propagation (LRP)~\cite{bach_2015_pixel} using 12 different rules.
In addition, we also evaluate explanations generated by the recently proposed B-Cos network adjustment~\cite{bohle_b-cos_2022}.

Similarly to Warnecke et al.~\cite{warnecke_evaluating_2020}, we evaluate the explainability methods based on four metrics: fidelity, sparsity, stability, and efficiency.
Since we only evaluate white-box methods that compute relevance vectors directly from the weights of a neural network, all explainability methods are complete in that they are able to compute non-degenerate explanations for every possible input.

In contrast to~\cite{warnecke_evaluating_2020}, we evaluate a total of 20 white-box explainability approaches (compared to the three evaluated by Warnecke et al.) and extend the fidelity and stability metrics to be more suitable for analyzing DGA classifiers.

Based on the four metrics, we select the top five techniques (\textit{b-cos}, \textit{deep\textunderscore taylor}, \textit{integrated\textunderscore gradients}, \textit{lrp.alpha\textunderscore 2\textunderscore beta\textunderscore 1}, and \textit{lrp.z\textunderscore plus}) for our bias investigation study in the next section.

% !TEX root = ../paper.tex
\section{Interpreting the Explanations}
\label{sec:bias_reveal}

\begin{figure}[!t]
	\centering
	\includegraphics[width=1.0\linewidth]{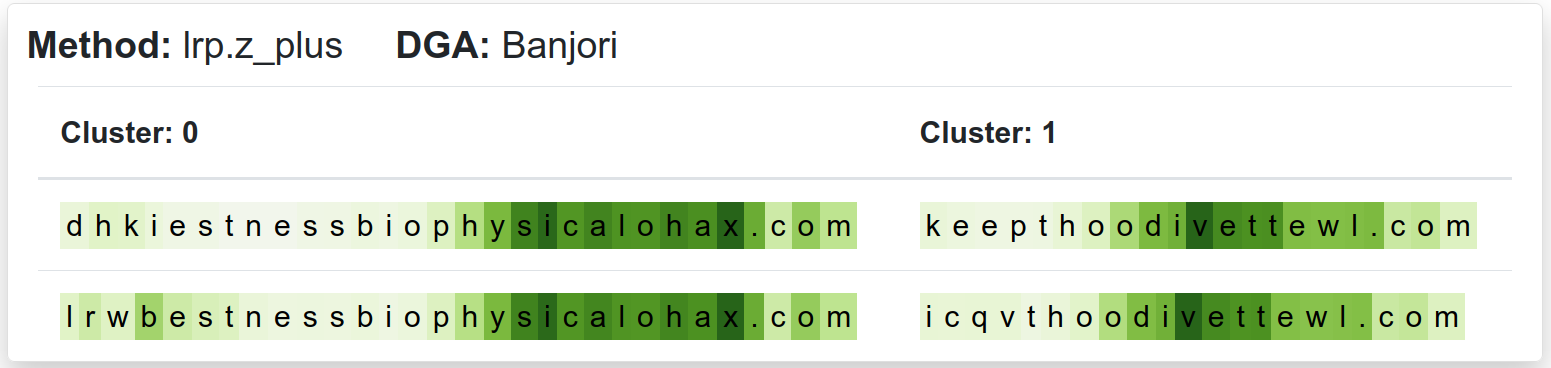}
	\caption{Example of clustering and visualization of domains.}
	\label{fig:example_clustering}
\end{figure}

Having decided on explainability methods, we can now examine the reasoning of the deep learning classifiers.
To this end, we use the classifiers trained during the four-fold cross validation on DS\textsubscript{mod} to predict all samples of DS\textsubscript{ex}, and then use all selected explainability methods to compute explanations.
Subsequently, for each method and class, we use DBSCAN~\cite{ester_dbscan_1996} to cluster the relevance vectors and group similar explanations together.
Finally, we manually review the clusters to identify potential features of the deep learning classifiers.
For each domain name and relevance vector, we visualize the importance of each character through heatmaps.
We encode positive contributions to the predicted label as green colors and negative contributions as red colors.
An example of the clustering and visualization of the relevance vectors generated by \textit{lrp.z\textunderscore plus} for the \textit{Banjori} DGA is shown in Fig.~\ref{fig:example_clustering}.\footnote{
Note that relevance vectors are not direct characteristics of individual inputs, but rather of the model that processes those inputs.
By clustering the relevance vectors, we can still find clusters similar to those in Fig.~\ref{fig:example_clustering}, but in this case it might be more appropriate to first compute clusters based on other features such as n-gram embeddings.
However, it is unclear what other features should be used to calculate such clusters (which brings us back to manual feature engineering) since, e.g., n-gram embeddings would not be useful for hex-based DGAs.
}

In the following we present our findings from this study.
We use the explainability methods to identify potential biases and then conduct various experiments to quantify the impact on classification.
While some of these biases may seem obvious and easily avoidable, they are present even in DGA detection approaches proposed at leading security conferences (e.g.,~\cite{schuppen_fanci_2018}).
Moreover, these biases are rooted on subtle flaws that are rife in security research and affect many other use-cases as well~\cite{arp_dos_2022}.

\subsection{Revealing Biases}
In this work, we mainly focus on the classification biases between the benign and the malicious class since the most severe danger in misclassification is that DGA-domains are wrongly labeled as benign.
If a certain proportion of samples is incorrectly assigned to a DGA by a multiclass classifier, this has less impact because the domains are still detected as malicious.
The main incentive for an adversary would be to exploit biases to force a detection system to classify DGA-domains as benign, allowing communication with botnets.
Therefore, we consider the threat model, which attempts to mask domains as if they were generated by another DGA, to be less reasonable.

In total, we identified five biases present in current state-of-the-art classifiers that provide a false sense of security, as they can be easily exploited to evade detection.\footnote{While we analyzed the ResNet-based classifier in detail, we verified that the identified biases are also exploitable in the LSTM-based~\cite{woodbridge_predicting_2016} and the CNN-based classifier~\cite{yu_character_2018}.}
Moreover, biases inherent in a classifier can affect the classifier's ability to detect yet unknown DGAs.

\subsubsection{Length Bias}
Across all explainability methods and across many clusters, dots included in a domain name are often calculated as particularly important for the classification.
We reckon that the dots themselves are not important in isolation, but that the deep learning classifiers infer the features of domain length and number of subdomains from it.

To assess the importance of this feature, we conduct the following experiment:
First, we chose the \textit{Qadars} DGA as it generates domains of a fixed length and is correctly attributed by M-ResNet most of the time (f1-score of 0.99400).
In detail, all domains generated by \textit{Qadars} match the following regular expression (regex): \mbox{\textasciicircum[a-z0-9]\{12,12\}\textbackslash.(com|net|org|top)\$}, i.e., \textit{Qadars} generates domains with a fixed length of 12, using only the characters a-z and 0-9, and finally adds a dot and one of four possible top-level domains (TLDs).
Then, we adapt the reimplementation of \textit{Qadars}\footnote{\url{https://github.com/baderj/domain\textunderscore generation\textunderscore algorithms}} to generate domains of all possible lengths.
Note that each domain name identifier can be a maximum of 63 characters long before it must be separated by a dot, and the full domain name can be a maximum of 253 characters long.
For each possible length and for each known seed (six in total), we generate at most 100 different domains, resulting in a dataset size of around 147,000 unique samples.
For each sample, we always fill in the highest level subdomain with characters before adding a dot.
Finally, we feed the generated domains into the M-ResNet classifier and observe the percentage of classifications assigned to \textit{Qadars}, any other DGA, and the benign class depending on the domain length.

\begin{figure}[!t]
	\centering
	\includegraphics[width=1.0\linewidth]{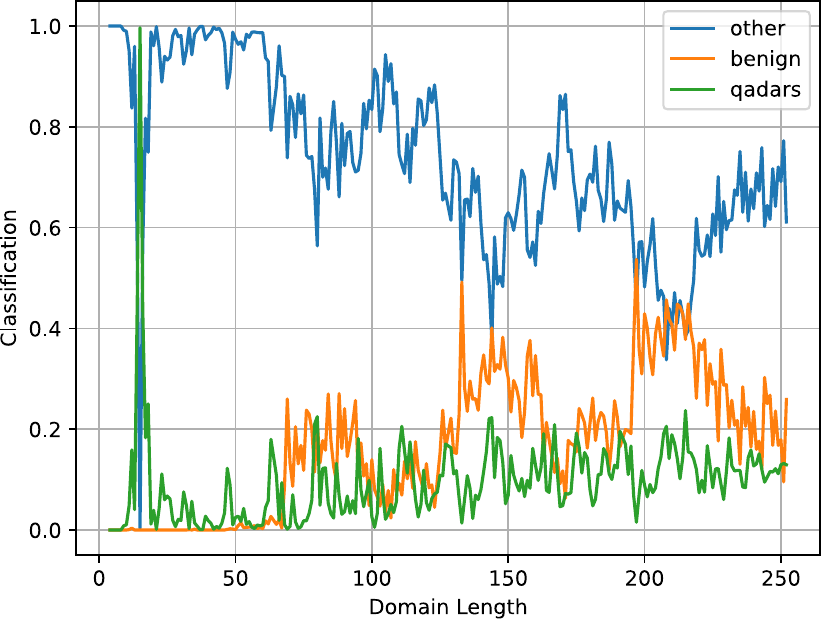}
	\caption{Results of the length bias experiment.}
	\label{fig:qadars_len}
\end{figure}

In Fig.~\ref{fig:qadars_len}, we display the results of this experiment.
The percentage of classifications assigned to \textit{Qadars} increases with domain length, peaking at the original domain length of 12, and then falls abruptly from there.
As the domain length increases, the percentage increases slightly because the classifier has more information to derive the correct prediction.
Most of the time, however, the classifier assigns the samples to different DGA classes.
The percentage of benign classifications increases rapidly from the length of 69, 133, and 197.
This is because at these lengths additional subdomains must be included to form a valid domain.
The more dots, the more benign classifications. Sometimes even more than 50\% of all classifications are assigned to the benign class.
After the dots are inserted, the benign classifications decrease with increasing domain length as more information generated by the DGA is available for prediction.

\begin{figure}[!t]
	\centering
	\includegraphics[width=1.0\linewidth]{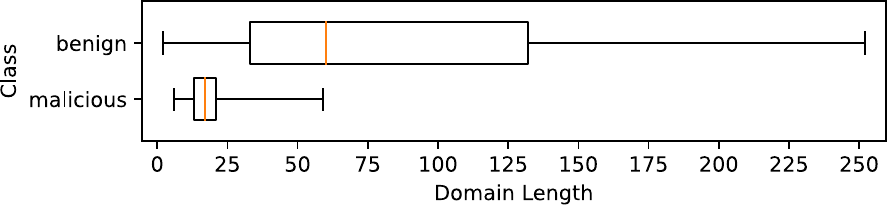}
	\caption{Box plots of domain length distribution within the training set for the benign and malicious classes.}
	\label{fig:statistics_len}
\end{figure}

Investigating the sample length distribution of the classifiers' training set illustrates the problem that with increasing length, more domains are classified as benign.
In Fig.~\ref{fig:statistics_len}, we display two box plots of the domain length distribution for the benign and malicious classes. 
The maximum domain length of a DGA-labeled sample within the training set is 59.
Thus, it is very likely that a classifier learns to assign a sample to the benign class with greater probability if it exceeds 59 in length.
Fortunately, this is not the only feature on which classification depends.
Since the domain length depends on the number of dots/subdomains, we examine this bias below.

\subsubsection{Number of Dots/Subdomains Bias}
As seen in the previous section, the number of dots/subdomains has a significant impact on the classification.
Looking at the number of dots contained in the training set separately for the benign and malicious classes, we can see that the benign class contains significantly more dots.
The average number of dots is 7.12, the median is 5, and the maximum is 35.
In comparison, the average for the malicious class is 1.08, the median is 1 and the maximum is 2.
In fact, only 19 DGAs generate domains with more than one dot and only two DGAs (\textit{Beebone} and \textit{Madmax}) have dots past their effective second-level domain (e2LD).
We refer to e2LD here because some DGAs use dynamic DNS services or public suffixes, which should not be counted as their generated second-level domain.

\subsubsection{www.~Bias}
In connection to the number of dots/subdomains bias we observed during our manual review of the relevance vector clusters for the benign class, that over all explainability methods, clusters have formed which highlight the importance of the ``www.'' prefix.
Examining the distribution of domains with the prefix ``www.'' within the training set, we find that the benign class contains 3,382 (0.00288\%) samples, while the malicious class contains only 183 (0.00016\%) samples.

To assess the impact of this bias, we perform the following experiment:
We take the four binary classifiers of the four-fold cross validation and all the malicious samples that the classifiers have correctly classified (true-positives).
Then we prepend the ``www.'' prefix to all true-positives and reevaluate the models on these samples.

On average over all folds, 434,916 (74.23\%) out of 585,907 true-positives became false-negatives, while only 150,991 were still correctly classified.
This shows that there is a huge bias regarding this prefix and malware authors could exploit this issue by simply prepending ``www.'' to their generated domains in order to evade detection of state-of-the-art classifiers.
Although, only a small fraction of all samples have the ``www.'' prefix, it can introduce bias into classification if the feature is sufficiently discriminatory.

\subsubsection{Top-Level Domain Bias}
Through our study, across all explainability methods and across multiple classes, we encountered multiple occurrences of clusters that, in combination with other features, highly value the top-level domain (TLD) as a significant feature.
To assess the impact of this feature, we make use of out-of-distribution (OOD) testing, as it was identified to be one of the most effective ways to reveal biases~\cite{geirhos_shortcut_2020}.
To this end, we perform a leave-one-group-out evaluation.
In detail, similarly to the four-fold cross validation, we train a classifier for every fold on the respective fold's training data of DS\textsubscript{mod}, except that we omit all samples of a particular class.
Then, we use the four trained classifiers to predict all samples of the left out class contained in DS\textsubscript{ex}.

As an example, we present the results obtained on the \textit{Mirai} DGA leave-one-group-out evaluation.
All samples generated by \textit{Mirai} use one of these three TLDs: \textit{online}, \textit{support}, and \textit{tech}.
In each fold all \textit{Mirai} samples that use the \textit{online} and \textit{tech} TLD are predicted to be malicious while all samples with the \textit{support} TLD are labeled as benign.

It seems that this is because the classifier tends to classify samples with never-seen TLDs into the benign class.
Omitting all \textit{Mirai} samples from training has the effect of removing all samples that use the \textit{support} TLD from the entire training set.
Although there appears to be enough information within the second-level domain to correctly assign a sample to the malicious class (as 100\% of all \textit{online} TLD samples are correctly assigned), the classifier is biased due to the unknown TLD to attribute the samples to the benign class.
Similar pictures emerge also for a variety of other DGAs.
Examination of the TLD distribution within the training set supports this statement.
There are 413 distinct TLDs in the benign data, of which 274 are unique to benign samples.
In comparison, there are only 258 different TLDs within the malicious labeled data, of which 115 are uniquely used by malicious samples.

On the other hand, all samples with the \textit{tech} TLD were also correctly labeled as malicious although this TLD was completely removed from the training data.
Since all \textit{support} TLD samples are misclassified and all samples use the same generation algorithm, it is unlikely that the information within the second-level domain was discriminatory enough for the \textit{tech} TLD samples.
Analyzing the calculated relevance vectors for these samples revealed that the classification is significantly influenced by the ``\textit{ch}'' suffix of the \textit{tech} TLD.
Looking at the \textit{ch} TLD distribution within the training data it becomes apparent why this is the case: there are 2063 \textit{ch} TLDs within the malicious samples and only 51 within the benign samples.

This bias investigation delivers two results: 
First, state-of-the-art classifiers heavily depend on the TLD, resulting in the fact that a malware author could simply change the TLD used to evade detection.
Second, it might be useful to encode the TLD as a one-hot encoded vector before inputting it to a classifier since it is rather a categorical feature.
In the case of the \textit{Mirai} evaluation, this was a stroke of luck for the defender site.
However, since the TLD can be freely chosen, an attacker could exploit this knowledge to evade detection.

\subsubsection{Validity/Diversity Bias}
During our study, we encountered several large benign clusters that contain domains that are invalid and therefore would not resolve (e.g. due to an invalid or missing TLD).
In fact, 7.64\% of all benign samples within the training set are invalid, while all malicious samples are valid.
An attacker has no incentive in generating invalid samples, as they would be useless for establishing connections between bots and their C2 server.
Thus, a classifier most likely learns the shortcut to distinguish domains based on their validity.
Although this is not a true bias, since invalid domains cannot be resolved and therefore assigned to the benign class, it does have an impact on the reported FPR of state-of-the-art classifiers as invalid samples are probably easier to classify.
While there is nothing wrong in calculating the FPR for the detection system which pre-filters invalid domains to the benign class, here the classifiers real true-negative rate (TNR) is artificially inflated.
Furthermore, including invalid samples in the training sets carries the additional risk of the classifier focusing on useless information and prevents the classifier from learning more complex features that might be useful in separating valid benign samples from malicious ones.

In addition, we found several benign clusters specific to the network in which the data was collected (e.g., domains including the official e2LD of the university).
Training and evaluating classifiers on this data could lead to misleadingly high results, as the classifiers may have only learned to separate network-specific domains from malicious ones, but they do not generalize between different networks.

% !TEX root = ../paper.tex
\section{Mitigating Biases}
\label{sec:bias_mitigation}
Now that we have identified several biases, we present strategies to mitigate them.
In addition, in various experiments, we measure the cost in terms of loss in classification performance for avoiding biases, since biases are nothing more than features that appear in the training data. For instance, biases such as the TLD are perfectly valid signals for the classifier to learn based on the underlying data distribution, since such features can be used to some extent to distinguish between benign and malicious samples. However, this is not desirable for features that can be easily modified by an attacker, as they can be exploited (e.g. by exchanging the TLD) to evade detection.
Finally, in a real-world study, we measure the true classification performance of DGA classifiers that are free of the identified biases, and evaluate whether a classifier generalizes to different networks and is time-robust.
In other words, here we evaluate whether a classifier is free from biases that might be introduced by artifacts in specific networks and at certain times.

\subsection{Mitigation Strategies}
In the following, we address the individual biases and suggest how to mitigate them.

\subsubsection{Number of Dots/Subdomains, www., and TLD Biases}
As demonstrated in the previous section, these biases can be easily exploited by an attacker to evade detection.
Adding the ``www.'' prefix to malicious domains converted around 75\% of true-positives into false-negatives, while selecting a TLD that was never seen by a classifier during training allows for complete bypass of detection.
Since the botmaster's authority over a domain starts with the e2LD and all other subdomains as well as the TLD can be freely selected, we suggest to perform the classification exclusively on the e2LD and to omit all other information.
Note that this does not open up any new attack vector, but may remove valuable features that could be used for classification, resulting in a decrease in overall classification performance.
Hence, in Section~\ref{sec:bias_mitigation_experiments}, we measure the trade-off between bias-reduced classification and performance.

\subsubsection{Validity/Diversity Bias}
Since invalid samples can be pre-filtered and assigned to the benign class, we choose to only train a classifier on valid domains, allowing the classifier to focus on task-relevant features.
As a result, the FPR of the classifier reported by us is likely to be larger than that reported by related work, since the classifier does not encounter easily classifiable invalid samples during testing.

Further, to mitigate the problem that a classifier only learns to separate network-specific domains from malicious ones, we focus on diverse data by training on unique e2LDs.
In doing so, we aim to train classifiers that generalize well between different networks.
Focusing solely on unique e2LDs has the effect that the underlying sample distribution changes fundamentally.
Training using this data will again increase the classifier's FPR since a e2LD occurs only once, either in the training or test set.
In contrast, in the state-of-the-art classification setting, a large proportion of unique domains with the same e2LD occur, which may be network-specific, such as domains that contain the university's official e2LD.
Once the classifier learns of a benign e2LD, samples with the same e2LD can be easily assigned to the benign class.

\subsubsection{Length Bias}
Focusing exclusively on valid and diverse e2LD already significantly equalizes the length distribution between benign and malicious samples and almost mitigates the bias.

\begin{figure}[!t]
	\centering
	\includegraphics[width=1.0\linewidth]{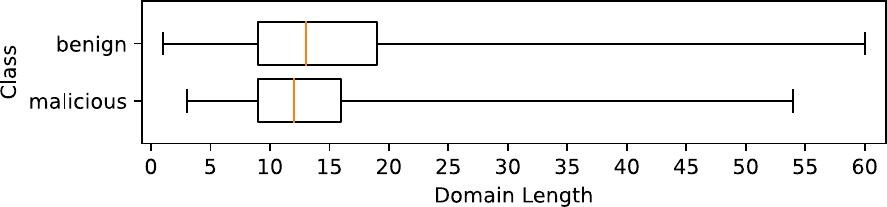}
	\caption{Box plots of unique and valid e2LD length distribution for the benign class and malicious samples.}
	\label{fig:statistics_len_new}
\end{figure}

In Fig.~\ref{fig:statistics_len_new}, we show two box plots of the unique and valid e2LD length distributions for the benign class and malicious samples.
In comparison to the sample length distributions in the state-of-the-art classification setting (cf. Fig.~\ref{fig:statistics_len}), the e2LD length distributions are much more similar.

Unfortunately, thereby the length bias cannot be fully mitigated.
The classifier will probably still tend to classify longer samples towards the benign class.
However, as we saw during the length bias experiment, longer samples contain more information that helps the classifier make the correct decision.
Thus, for an adversary, increasing the domain length is more of a trade-off between exploiting length bias and providing too much information to the classifier.

Note, reducing the domain length of input samples to mitigate this bias is not a viable option, as this opens up a new attack vector where an attacker can hide features that would have sorted a domain into the malicious class.

On the other hand, it is possible to generate additional artificial domains by adapting publicly available reimplementations of DGAs (similar to the length bias experiment) to balance the length distributions and thus mitigate the bias completely.
However, this may require oversampling of benign data and care must be taken to ensure that this does not affect classification performance on clean data.
Since the focus on valid and diverse e2LD almost evens out the distributions, we decided against it.

\subsection{Bias Mitigation Experiments}
\label{sec:bias_mitigation_experiments}
In the following, we measure the cost in terms of loss in classification performance for avoiding biases.
We expect classification performance to deteriorate because biases are nothing more than features based on the underlying distribution of the training data.
All experiments are similar to the four-fold cross validation performed in Section~\ref{sec:sota_result_reproduction}, except that here we focus on diverse data.
To this end, we first map all fully qualified domain names (FQDNs) to their e2LDs.
We then randomly sample the e2LDs and then select exactly one sample per unique e2LD for each evaluation scenario. 

For binary and multiclass classification, we examine four scenarios each: classification on valid and diverse FQDNs, on FQDNs without TLDs (no TLDs), on FQDNs without subdomains (e2LDs + TLDs), and exclusively on e2LDs. 

\begin{table}
%	\scriptsize
	\caption{Results of the bias mitigation experiments.}
	\label{tab:bias_mitigation_experiments}
	\centering
	\resizebox{\linewidth}{!}{
		\begin{tabular}{lcccc}
			\hline \Tstrut
			\textbf{Setting} & \textbf{Scenario} & \textbf{ACC} & \textbf{TPR} & \textbf{FPR} \\
			\hline \Tstrut
			Binary & state-of-the-art & 0.99864 & 0.99982 & 0.00255 \\
			\hline \Tstrut
			Binary & valid/diverse FQDNs & 0.96916 & 0.98068 & 0.04243 \\
			Binary & no TLDs & 0.95416 & 0.97949 & 0.07132 \\
			Binary & e2LDs + TLDs & 0.93076 & 0.93704 & 0.07555 \\
			Binary & e2LDs & 0.89139 & 0.88824 & 0.10544 \\
			\toprule
			\toprule
			\textbf{Setting} & \textbf{Scenario} & \textbf{F1-Score} & \textbf{Precision} & \textbf{Recall} \\
			\hline \Tstrut
			Multiclass & state-of-the-art & 0.78682 & 0.80058 & 0.79690 \\
			\hline \Tstrut
			Multiclass & valid/diverse FQDNs & 0.77878 & 0.79816 & 0.78209 \\
			Multiclass & no TLDs & 0.60126 & 0.62220 & 0.62016 \\
			Multiclass & e2LDs + TLDs & 0.77113 & 0.79057 & 0.77588 \\
			Multiclass & e2LDs & 0.58836 & 0.61533 & 0.60968 \\
		\end{tabular}
	}
\end{table}

In the upper part of Table~\ref{tab:bias_mitigation_experiments}, we present the results for the binary setting while the lower part of the table displays the results for the multiclass setting.  
For convenience we also show the performance of the classifiers in the state-of-the-art classification setting from Section~\ref{sec:sota_result_reproduction}. 

As suspected, when only valid and diverse samples are used, the performance of the binary classifier is significantly worse, especially with respect to the FPR.
Removing the TLDs from the FQDNs has less of an impact on performance than removing all subdomains after the e2LD.
However, in both scenarios the loss in performance is tremendous, increasing the FPR to about 7.1\% - 7.6\%.
Classification solely on the e2LD delivers the worst results reaching a 89.1\% TPR @ 10.5\% FPR for the decision threshold of 0.5.
Examining the individual TPRs for each DGA, we find that the rate drops significantly for some DGAs, while for others it remains high, even reaching 100\%.
Although the average TPR drops significantly compared to the state-of-the-art setting, we expect that most DGAs could still be detected as they query multiple domains before finally resolving a registered domain.
Provided that a decision is not made on the basis of a single query.
Only the DGAs \textit{Redyms} and \textit{Ud3} would be completely missed as for these DGAs the TPRs are zero over all four folds.

In the multiclass setting, classification performance is not affected as much when trained on valid and diverse FQDNs.
This is because focusing on these samples mainly affects the benign class and a few DGA classes that have a small sample size and generate FQDNs that map to the same e2LD (e.g., they generate domains with the same e2LD but with different TLDs).
However, most DGAs are not affected by this.
In contrast to the binary setting, here the TLDs are more relevant for classification than the subdomains after the e2LD.
If only the e2LDs are used for classification, the performance deteriorates drastically (mainly because of the missing TLDs).
Removing all subdomains after the e2LD affects only two DGAs: \textit{Beebone} and \textit{Madmax}.
However, when the subdomains are removed, there is still enough information in their domain names to classify them correctly most of the time.
\textit{Beebone's} f1-score drops slightly from 97.7\% to 95.7\%, and \textit{Madmax's} from 74.9\% to 60.2\%.

In summary, the TLD is vital for the multiclass classification.
In the binary setting, classifying exclusively e2LD is as bias-free as possible but the achieved performance does not seem to be acceptable.
However, the effective TPR@FPR operation point of a detection system that pre-filters invalid samples and classifies all input samples regardless of the uniqueness of their e2LD can still be acceptable.
In the next section, we get to the bottom~of~this~question.

\subsection{Real-World Study}
\label{sec:rw_study}
In this section, we perform a real-world study to assess the true performance of bias-reduced DGA binary classification.
In this context, we evaluate whether the classifiers generalize between different networks and are time-robust.
Simultaneously, we enforce that the evaluation is free of experimental biases.
In the following, we refer to classifiers that mitigate the identified biases as bias-reduced classifiers.

To this end, we train a classifier using the real-world benign e2LDs from the university network recorded from mid-October 2017 to mid-November 2017, as well as DGArchive data that was available until the end of the recording period.
In detail, DGArchive contains approximately 53 million unique domains generated by 85 different DGAs up to this point in time.
Training a classifier using a dataset which is similar to DS\textsubscript{mod}, but with the constraint that the malicious samples are from the same time window as the benign samples, mitigates one of the two experimental temporal biases included in the state-of-the-art classification setting.
To mitigate the second experimental temporal bias, that requires that all training samples are strictly temporally precedent to the testing ones, we evaluate the classifier on approximately 311 million benign e2LDs captured in the company network in April 2019~(cf. Section~\ref{sec:data}) and DGA-domains from DGArchive that were generated by DGAs in April 2019.
Within April 2019, 46 DGAs (four of which were unknown at the time of the training) generated approximately 1.2 million domains.
In this way, we eliminate the experimental temporal biases, and can guarantee that the benign samples come from different networks and that the time interval between the occurrence of the training and the test samples is about 17 months.

To eliminate the experimental spatial bias, it is required to approximate the true ratio of benign to malicious samples in the test data.
Since the true sample distribution is unknown, we conduct two experiments to estimate the true detection performance of bias-reduced DGA binary classification.

First, we evaluate the classifier using all 311 million benign e2LDs and gradually increase the amount of included malicious test samples generated in April 2019 from 1\% to 100\% for each DGA.
Thereby, the ratios between the domains generated by the different DGAs follow the true distribution.

\begin{figure}[!t]
	\centering
	\includegraphics[width=1.0\linewidth]{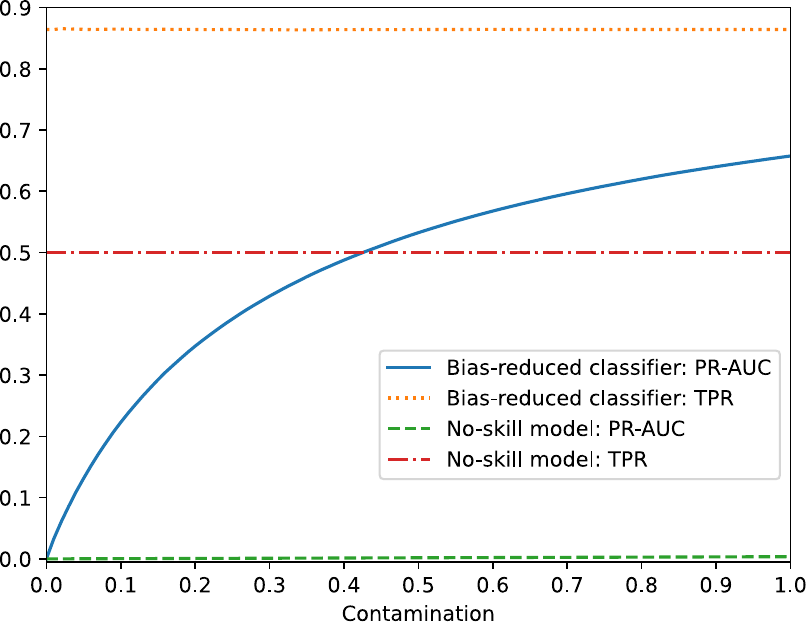}
	\caption{PR-AUC and TPR scores, depending on the percentage of April 2019 DGA-domains included in the test set.}
	\label{fig:rw_contamination}
\end{figure}

In the following, we report the obtained results of the classifier that first checks whether a sample is invalid.
If it is invalid, the sample is ignored.
Otherwise, it is evaluated by the classifier.
In Fig.~\ref{fig:rw_contamination}, we display the Precision-Recall Area Under the Curve (PR-AUC) and the TPR for the decision threshold of 0.5 of the bias-reduced classifier depending on the contamination of the test set (i.e., the relative amount of included malicious test samples from April 2019).
In addition, we present both metrics for the no-skill model which classifies all samples uniformly at random.
At the beginning, the PR-AUC value of the bias-reduced classifier increases faster than towards the end, reaching a value of about 0.66 at 100\% contamination.
A steeper initial slope indicates a classifier whose precision is less affected by very small base rates of malicious samples.
The achieved TPRs are nearly stable for all ratios of benign to malicious samples.
The bias-reduced classifier is far better than the no-skill model, whose precision always corresponds to the proportion of malicious samples in the test set.
This experiment quantifies the natural impact of different base rates of malicious samples on the classifier's precision.
Note that the benign data heavily overshadows the malicious data even when we include 100\% of all DGA-domains from April 2019.
Here, the relative percentage of malicious samples varies between 0.00362\% and 0.35998\%, which means that in the worst case, 99.64002\% of the test data is still from the benign class.

As it is unclear, how many DGAs are present in a real-world network, we additional conduct a second experiment to estimate the worst-case classification performance.
Here, for each DGA, we evaluate the classifier using all malicious samples generated in April 2019 of that particular DGA and all 311 million benign e2LDs.
In total, we thus evaluate the classifier using 46 test sets, since there are 46 DGAs that generate at least one domain in April 2019.

\begin{figure}[!t]
	\centering
	\includegraphics[width=1.0\linewidth]{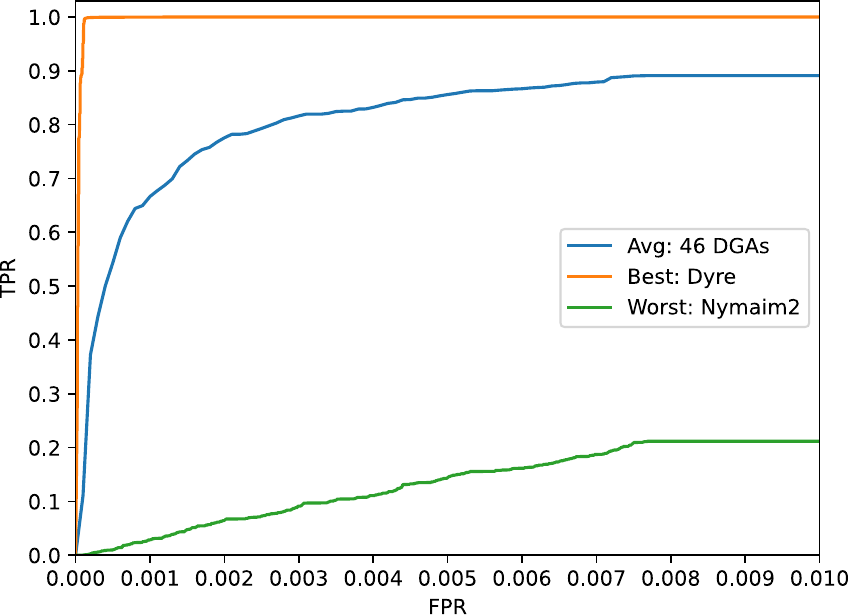}
	\caption{Estimated worst-case ROC curve averaged over all 46 evaluation runs of the bias-reduced classifier including the ROC curves for the best-detected DGA (\textit{Dyre}) and the worst-detected DGA (\textit{Nymaim2}).}
	\label{fig:rw_roc_dgas_avg}
\end{figure}

On average the bias-reduced classifier achieves a TPR of 0.85735 at a FPR of 0.00506 for the decision threshold of 0.5.
In Fig.~\ref{fig:rw_roc_dgas_avg}, we display the receiver operating characteristic (ROC) curve averaged over all evaluation runs for the FPR range of $[0,0.01]$.
In addition, we also show the ROC curves for the best-detected DGA (\textit{Dyre}) and the worst-detected DGA (\textit{Nymaim2}).

We argue that the classifier is remarkable time-robust and generalizes well to different networks.
The temporal and spatial changes in data distribution have increased the FPR compared to the state-of-the-art setting at the decision threshold of 0.5.
However, this was to be expected as the distribution of benign samples naturally varies between networks, at least to some degree.
Moreover, the classifier is able to achieve a slightly lower TPR as the bias-reduced e2LD classifiers from the previous section.
Surprisingly, for three of the four DGAs that were unknown at the time of training (\textit{Ccleaner}, \textit{Tinynuke}, \textit{Wd}), the bias-reduced classifier is able to correctly classify 100\% of all generated samples.
Only the \textit{Nymaim2} DGA is detected worse with a TPR of 14.84\%, which is the main reason for the slightly lower average TPR compared to the bias-reduced e2LD classifiers from the previous section.\footnote{
We additionally evaluated the four e2LD classifiers from the previous section against the 311 million benign NXDs and all DGA-domains from DS\textsubscript{ex} (which are completely disjoint with the training samples) to evaluate the performance using all 106 known DGAs.
Thereby, we arrive at very similar results.
We present the corresponding ROC curves in Appendix~\ref{sec:appendix}.
Note that this of course reintroduces experimental temporal bias.
}

At a fixed FPR of 0.008 the bias-reduced classifier achieves a TPR of about 89\%.
In practice, it might be advantageous to set the threshold to a lower fixed FPR value.
Setting the FPR at 0.001 to 0.002 would still allow an approximate detection rate of about 67\% to 78\%.
However, how useful this is and to what extent the base-rate fallacy limits the use of the classifier in practice depends on what is done with the classification results.
Context-less DGA detection was never intended for single-domain based decision-making.
This evaluation assessed the true performance of bias-reduced DGA classifiers and demonstrated the limits of what is possible without contextual information.

% !TEX root = ../paper.tex
\section{Bias-reduced DGA Classification}
\label{sec:new_system}
In this section, we use the insights gained from the bias mitigation and the real-world study to propose a classification system that (1) is as bias-free as possible and (2) does not miss entire DGA families.
Further, we propose an approach to improve visualization support to increase trust in and transparency of detection methods and facilitate decision-making.

\subsection{Bias-reduced DGA Classification System}
As previous evaluations have shown, bias can be easily exploited to evade detection.
Focusing exclusively on e2LD helps mitigate most identified biases.
However, this causes the classifier to lose the ability to recognize specific DGA families as a whole.
In the case of multiclass classification, we have seen that the classification relies heavily on information outside of the e2LD to correctly assign domains of multiple classes.

\begin{figure}[!t]
	\centering
	\includegraphics[width=1.0\linewidth]{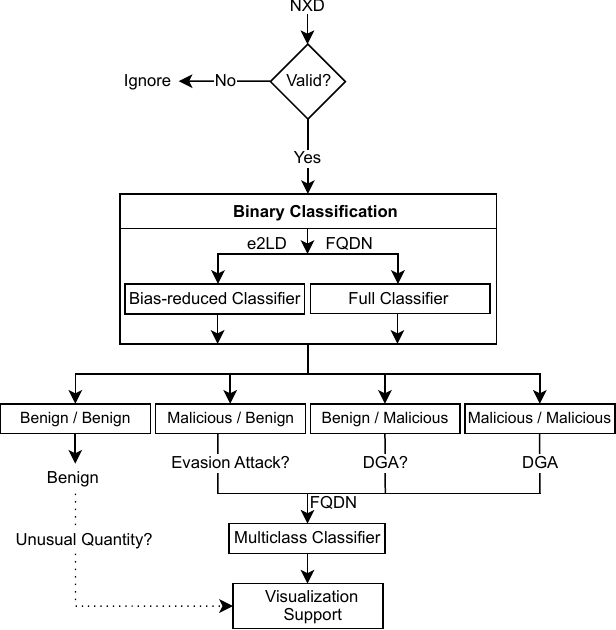}
	\caption{Bias-reduced DGA classification architecture.}
	\label{fig:system}
\end{figure}

In the following, we present a detection system that counteracts these issues.
In Fig.~\ref{fig:system}, we visualize the system's architecture.
In the first step, the detection system evaluates whether the entered NXD is invalid or not.
If it is invalid, it is ignored, otherwise the input sample is passed to the binary classification step. 
Here, two classifiers work in parallel: a bias-reduced classifier that classifies the e2LD of the input sample, and a full classifier that uses the FQDN.
This classification step can lead to four possible outcomes:
First, both classifiers agree on the benign label, so the detection system also outputs benign.
Second, the bias-reduced classifier outputs malicious while the full classifier predicts benign.
This is an indication that an attacker might try to exploit biases to evade detection.
Third, the bias-reduced classifier predicts benign and the full classifier malicious.
This suggests that the features outside the e2LD may be indispensable to detect the DGAs that the bias-reduced classifier would miss.
And fourth, both classifiers agree on the malicious label indicating that the input sample is very likely DGA-generated.
Regardless of the results, the input sample can be passed to a multiclass classifier trained on FQDNs to associate the sample with the DGA that most likely generated it.
Finally, we propose to pass the input sample associated with the classification results to a visualization system to understand the classifier's reasoning and to support the decision-making process.

Using this detection system, we achieve bias-reduced DGA detection and do not miss entire DGA families.

\subsection{Visualization Support}
The proposed detection system gets the most out of context-less and bias-reduced DGA classification.
In order to facilitate decision-making and to better understand the reasoning of a classifier we propose a visualization system. 
In this work, we demonstrated the limits of context-less classification and showed that decision-making based on the classification result of a single query is practically insufficient.
To make a decision based on multiple classification results, the minimum information required is the mapping between the host and the queried domains.
While this information may not be available to a CaaS provider, the network operator that uses the service most likely has this knowledge.
In the following, we only use this additional knowledge to facilitate the work of SOC~analysts.

\begin{figure}[!t]
	\centering
	\includegraphics[width=1.0\linewidth]{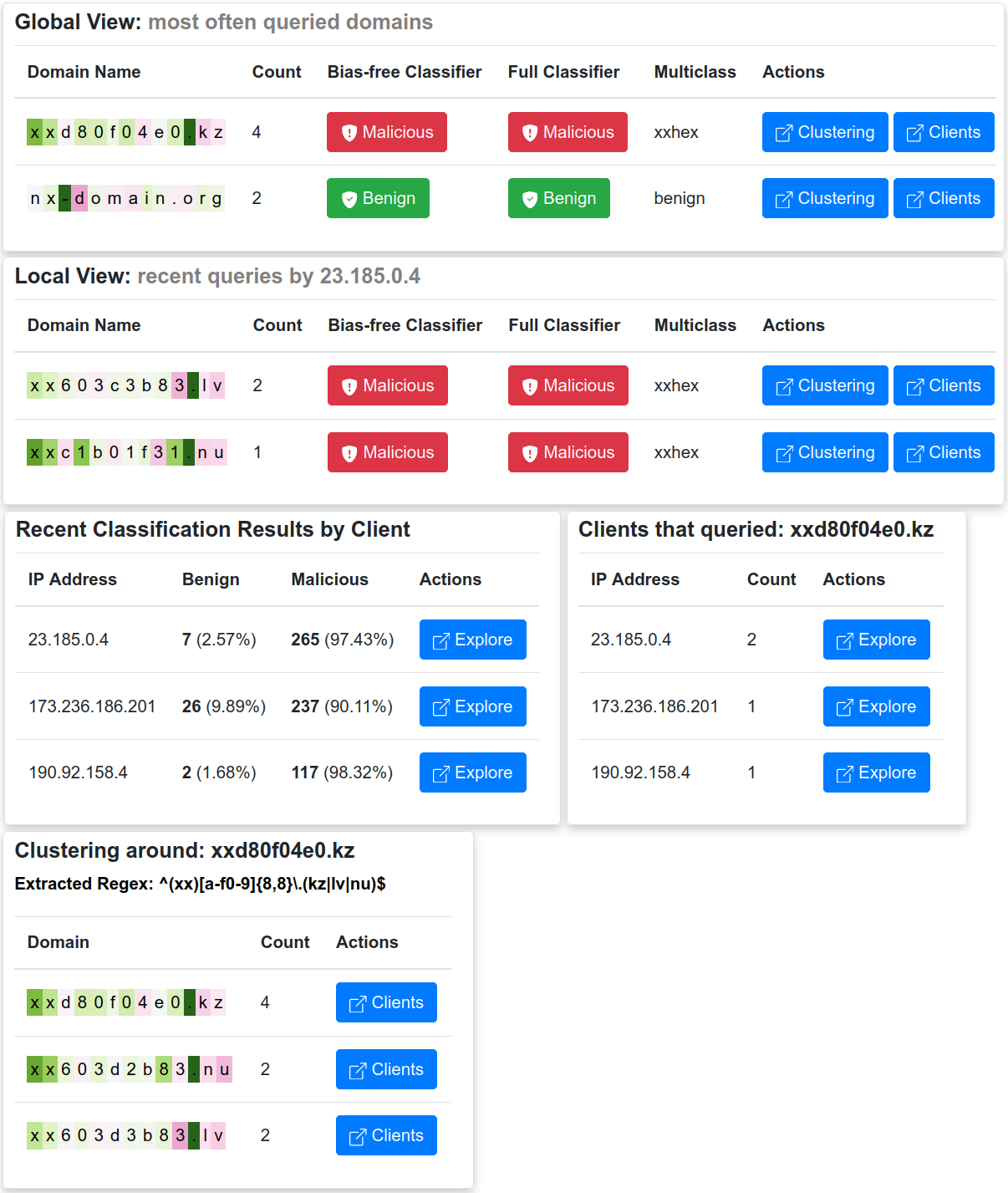}
	\caption{Different views of the proposed visualization system that help in decision-making.}
	\label{fig:visualization_system}
\end{figure}

Fig.~\ref{fig:visualization_system} shows the different views of the proposed visualization system based on mock data.
Two main view groups summarize the classification results: the global and the local views.
Both contain the queried domain names, in which the relevance of each character to the prediction is highlighted using a heatmap.
In this example, we used \textit{integrated\textunderscore gradients} to compute the relevance vectors for the predictions of the multiclass model.
However, any other explainability method can be chosen.
In addition, we display the total amount of times the domain was queried as well as the classification results from the bias-reduced, full binary, and multiclass classifier.
The global view summarizes all classification results for the entire network and allows finding multiple hosts infected with the same malware.
The local view summarizes the results for a single host and allows targeted analysis of all queries performed by that host.
Local views can be accessed through the \textit{Recent Classification Results by Client} view, which displays the total and relative number of domains classified as benign or malicious per host.
From both, the global and the local view, it is possible to analyze how often and which hosts queried a particular domain.
Additionally, for each domain, it is possible to analyze the clusters in which the relevance vector falls and to extract a simple regex that fits all samples within the cluster.
In this way, it may be possible to identify multiple hosts infected with the same malware.

% !TEX root = ../paper.tex
\section{Additional Utilization of the Knowledge Gained}
\label{sec:improve_sota}
As a secondary contribution, we use the knowledge gained in the previous evaluations to improve the state-of-the-art deep learning and feature-based multiclass classifiers in terms of classification performance and efficiency.
In this section, we therefore take a step back from improving the generalization of classifiers by removing classification biases and briefly turn our attention to improving the performance and efficiency of the classifiers themselves.

\subsection{Improving M-ResNet}
In this work, we mainly improved the binary classifier B-ResNet by mitigating identified biases.
Now we also take a closer look at the multiclass classifier M-ResNet.
In Section~\ref{sec:bias_reveal}, we noted that the classifier does not use the TLD as a standalone feature, but also derives additional features from the character distribution.
Since the TLD can be freely chosen by the adversary and the TLD is more of a categorical feature, we adapt the M-ResNet model to classify a domain by using the one-hot encoded vector representation of the TLD instead of the character-wise encoding.
Thereby, we aim to improve classification performance by allowing the classifier to focus on the more important part of the FQDN.
Furthermore, this has the effect that other implicit features, such as domain length, are no longer affected by the chosen TLD.

We evaluated this model using a four-fold cross validation on DS\textsubscript{mod} but could not measure any significant improvement.
As could be seen in the relevance vector cluster analysis, the original model appears to have a large enough capacity to learn the correct extraction of the TLD from the characters.
Furthermore, the characters within the TLD do not appear to significantly affect the multiclass classifier.
Since overparameterization has been associated with a higher susceptibility to learning spurious correlations~\cite{sagawa_investigation_2020}, we attempt to iteratively reduce the complexity of the adapted model.
As a result, we were able to successfully remove the last four residual blocks and reduce the number of trainable parameters by 35.5\% without affecting classification performance (f1-score of 0.78691).
Thereby, we additionally improved the model's carbon footprint and reduced the required time for training and inference.

\subsection{Improving EXPLAIN}
Now we try to improve the feature-based multiclass classifier EXPLAIN by using knowledge extracted by explainability methods applied on M-ResNet.
To this end, we cluster relevance vectors for samples which are correctly classified by M-ResNet but incorrectly by EXPLAIN, targeting the identification of features that are missing in EXPLAIN.

We attribute the performance difference between both classifiers to four findings: (1) ResNet seems to handle imbalanced data and class weighting better, (2) for some DGAs, M-ResNet is simply better at guessing, (3) M-ResNet is able to learn complex features through a series of non-linear transformations that are not easily understood by a human, and (4) both classifier converge to different local optima and thus tend to assign similar samples to either one or the other class.

\subsubsection{Imbalanced Data}
Investigating the relevance vector clusters for the \textit{Redyms} DGA, it is immediately apparent that for M-ResNet, the ``-'' character is useful for the correct classification.
Although, the feature that counts the ``-'' character is defined in EXPLAIN's source code, it was not selected during the feature selection process.
We reckon, that this is because the feature is only important for a few classes but other features are important for a much higher number of classes which resulted in lower importance score during the feature selection process.
This problem could be the reason why several classes are recognized worse by EXPLAIN, and suggest that M-ResNet might be better with imbalanced data and class weighting in general.
In contrast to EXPLAIN's feature selection step, we assume that M-ResNet does not completely remove self-learned features, but fine-tunes the importance by adjusting the weights.
Adding the ``-''-feature to EXPLAINs feature set improves the f1-score for the \textit{Redyms} DGA by 53.15\% and brings the detection rate to a level similar to that of M-ResNet.

\subsubsection{Random Guessing}
EXPLAIN mostly confuses the samples of \textit{Ud4} with \textit{Dmsniff}.
Analysis of all samples from both classes revealed that both DGAs generate 100\% identical domains, so they are most likely the same DGA.
Upon inquiry to DGArchive this was confirmed and in the future the feed of \textit{Ud4} will be discontinued.
Here, M-ResNet is just better at guessing (by an f1-score of 16.48\%).

\subsubsection{Complex Features}
We cannot exclude the possibility that M-ResNet is able to learn complex features through a series of non-linear transformations that are not easily understood by a human.
For instance, related work~\cite{drichel_analyzing_2020} suggests that the ResNet classifier may be able to distinguish, at least to some degree, between underlying pseudo-random number generators.
To improve EXPLAIN, we adapt the features related to randomness tests and add all of them to the final feature set.
In detail, we adapt the 14 randomness tests from ~\cite{rukhin_statistical_2001} to include the final p-values used for the decision of whether a certain randomness test is passed instead of only the result of the test.
Reevaluating the model with all additional features, we could measure a small improvement of 0.783\% in f1-score.

\subsubsection{Different Optima}
Most other DGAs that are confused by EXPLAIN generate similar domains, and often all domains match the same regexes.
EXPLAIN is significantly better (> 10\% in f1-score) than M-ResNet in four DGAs, whereas M-ResNet is also significantly better in four other DGAs.
We reckon that both models converge to different local optima and thus tend to assign similar samples to either one or the other class.

\subsubsection{Overall Results}
We were able to improve EXPLAIN from an f1-score of 0.76733 to 0.77516 by adding additional features to EXPLAINs feature set, bringing it closer to the performance of deep learning classifiers such as M-ResNet.

% !TEX root = ../paper.tex
\section{Other Related Work}
\label{sec:related_work}

We already discussed related work on DGA detection in Section~\ref{sec:dga_detection}.
Consequently, we focus here on related work on explainability and bias learning prevention.

For the DGA detection use-case, there are only a few works that partially address the explainability of detection systems.
Drichel et al.~\cite{drichel_first_2021} proposed the multiclass classifier EXPLAIN as a feature-based alternative to deep learning-based classifiers.
While feature-based approaches often seem inherently explainable, it is often not easy to interpret their predictions.
For instance, EXPLAIN's predictions are based on the majority vote of 360 decision trees with a maximum depth of 43 and a random mixture of 76 features that include several statistical features that are difficult for a human to analyze.
The authors of~\cite{piras_expaining_2022} also adopt a feature-based RF classifier based on the EXPOSURE system~\cite{bilge_exposure_2014} and mainly use SHAP~\cite{lundberg_unified_2017} to derive explanations.
However, their approach relies heavily on extensive tracking of DNS traffic and is unable to derive explanations in the multiclass classification setting.

None of these works investigate biases inherent in detection methods.
To the best of our knowledge, this is the first work to critically analyze the features used, focusing on their limitations and unintended consequences for the DGA use-case.

In addition, related work~\cite{geirhos_shortcut_2020} has identified several general measures to mitigate bias learning that can also be applied here.
Changing the loss function~\cite{jacobsen_excessive_2019} and adding regularization terms~\cite{heinze_conditional_2017,arjovsky_invariant_2019} can force a classifier to learn more complex features instead of focusing on simple biases.
Also, the learning rate of the optimizer can be adjusted to make the classifier learn either simpler or more complex features~\cite{li_towards_2019,arpit_closer_2017}.

Somewhat related is the issue of adversarial attacks and the robustness of classifiers.
Here, semantic gaps in the data create blind spots in classifiers which make them susceptible to small input perturbations that lead to misclassifications.
Adversarial training can be used to prevent such classification shortcuts~\cite{madry_towards_2017}.
In context of DGA detection, several works deal with this topic \mbox{(e.g.~\cite{drichel_analyzing_2020,spooren_detection_2019,peck_charbot_2019,anderson_deepdga_2016})}.

% !TEX root = ../paper.tex
\section{Conclusion}
\label{sec:conclusion}

In this work, we showed how XAI methods can be used to debug, improve understanding, and enhance state-of-the-art DGA classifiers.
To this end, we performed a comparative evaluation of different explainability methods and used the best ones to explain the predictions of the deep learning classifiers.
Thereby, we identified biases present in state-of-the-art classifiers that can be easily exploited by an adversary to bypass detection.
To solve these issues we proposed a bias-reduced classification system that mitigates the biases, achieves state-of-the-art detection performance, generalizes well between different networks, and is time-robust.
In this context, we measured the true performance of state-of-the-art DGA classifiers, showed the limits of context-less DGA binary classification, and proposed a visualization system that facilitates decision-making and helps to understand the reasoning of deep learning classifiers.
Finally, we used the knowledge gained from our study to improve the state-of-the-art deep learning as well as feature-based approaches for DGA multiclass classification.

In future work, the usefulness of the visualization system needs to be evaluated, preferably in an operational environment.
A promising future research direction is the combination of context-less and context-aware systems to further enhance detection and decision-making.

\section*{Availability}
We make the source code of the machine learning models publicly available\footnote{\url{https://gitlab.com/rwth-itsec/explainability-analyzed-dga-models}} to encourage replication studies and facilitate future work.

%%
%% The acknowledgments section is defined using the "acks" environment
%% (and NOT an unnumbered section). This ensures the proper
%% identification of the section in the article metadata, and the
%% consistent spelling of the heading.
\begin{acks}
The authors would like to thank Daniel Plohmann, Simon Ofner, and the Cyber Analysis \& Defense department of Fraunhofer FKIE for granting us access to DGArchive as well as Siemens AG and Jens Hektor from the IT Center of RWTH Aachen University for providing NXD data.
\end{acks}

%%
%% The next two lines define the bibliography style to be used, and
%% the bibliography file.
\bibliographystyle{ACM-Reference-Format}
\bibliography{bibliography}

%%
%% If your work has an appendix, this is the place to put it.
\appendix
% !TEX root = ../paper.tex
\section{Evaluating Explainability Methods}
\label{sec:evaluating_explainability_methods}
We evaluate the explainability methods using four metrics: fidelity, sparsity, stability, and efficiency following~\cite{warnecke_evaluating_2020}.
Since we only evaluate white-box methods that compute relevance vectors directly from the weights of a neural network, all explainability methods are complete in that they are able to compute non-degenerate explanations for every possible input.

To evaluate the explainability methods we use the four classifiers trained on DS\textsubscript{mod} during our results reproduction study and predict all samples from DS\textsubscript{ex}.
For each metric, we average the results across all classifiers.

\begin{table*}
	\tiny
	\caption{Results of the evaluated explainability methods averaged over the explanations derived for four classifiers.}
	\label{tab:rlv_results}
	\centering
	\resizebox{\linewidth}{!}{
		\begin{tabular}{lccccc}
			\hline \Tstrut
			\textbf{Method} & \textbf{Fidelity: removed / replaced} & \textbf{Sparsity}& $\mathbf{Sparsity * (1-Fidelity)}$ & \textbf{Stability} & \textbf{Efficiency} \\
			\hline \Tstrut
			\textbf{b-cos} & 0.13253 / \textbf{0.20002} & 0.64770 & 0.56186 / 0.51815 & 0.17335 & 0.00988 \\
			\hline \Tstrut
			deconvnet & 0.25730 / 0.40557 & 0.57508 & 0.42712 / 0.34185 & 0.22411 & 0.00022 \\
			\textbf{deep\textunderscore taylor} & 0.13146 / 0.30295 & 0.66335 & 0.57615 / 0.46239 & \textbf{0.08051} & 0.00054 \\
			gradient & 0.16483 / 0.28481 & 0.68988 & 0.57617 / 0.49340 & 0.28186 & 0.00022 \\
			guided\textunderscore backprop & 0.15936 / 0.24556 & 0.64526 & 0.54243 / 0.48681 & 0.15865 & 0.00023 \\
			input\textunderscore t\textunderscore gradient & 0.14022 / 0.28705 & 0.73299 & 0.63020 / 0.52258 & 0.21487 & 0.00022 \\
			\textbf{integrated\textunderscore gradients} & \textbf{0.12779} / 0.25919 & 0.75136 & \textbf{0.65534 / 0.55661} & 0.18180 &  0.00924 \\
			lrp.alpha\textunderscore 1\textunderscore beta\textunderscore 0 & 0.16233 / 0.23764 & 0.60709 & 0.50854 / 0.46283 & 0.16565 & 0.00064 \\
			\textbf{lrp.alpha\textunderscore 2\textunderscore beta\textunderscore 1} & 0.19819 / 0.39214 & \textbf{0.79095} & 0.63420 / 0.48079 & 0.22698 & 0.00092 \\
			lrp.alpha\textunderscore 2\textunderscore beta\textunderscore 1\textunderscore IB & 0.16165 / 0.30515 & 0.75296 & 0.63124 / 0.52320 & 0.21837 & 0.00087 \\
			lrp.flat & 0.19219 / 0.37522 & 0.68880 & 0.55642 / 0.43035 & 0.27051 & 0.00035 \\
			lrp.sequential\textunderscore preset\textunderscore a & 0.16374 / 0.23851 & 0.61205 & 0.51183 / 0.46607 & 0.18329 & 0.00062 \\
			lrp.sequential\textunderscore preset\textunderscore a\textunderscore flat & 0.16845 / 0.24264 & 0.57311 & 0.47657 / 0.43405 & 0.19896 & 0.00060 \\
			lrp.sequential\textunderscore preset\textunderscore b & 0.18008 / 0.35645 & 0.78989 & 0.64765 / 0.50834 & 0.22260 & 0.00091 \\
			lrp.sequential\textunderscore preset\textunderscore b\textunderscore flat & 0.20016 / 0.36106 & 0.74873 & 0.59886 / 0.47839 & 0.25922 & 0.00088 \\
			lrp.w\textunderscore square & 0.19254 / 0.37443 & 0.68820 & 0.55570 / 0.43052 & 0.27479 & 0.00035 \\
			lrp.z & 0.14023 / 0.28705 & 0.73299 & 0.63020 / 0.52258 & 0.21487 & 0.00034 \\
			\textbf{lrp.z\textunderscore plus} & 0.15644 / \textbf{0.22766} & 0.59925 & 0.50550 / 0.46283 & 0.10362 & 0.00056 \\
			lrp.z\textunderscore plus\textunderscore fast & 0.16604 / 0.29094 & 0.75430 & 0.62905 / 0.53484 & 0.23621 & 0.00034 \\
			smoothgrad & 0.20429 / 0.44121 & 0.67668 & 0.53844 / 0.37813 & 0.32758 & 0.00926 \\
		\end{tabular}
	}
\end{table*}

\subsection{Fidelity}
The first evaluation criterion is fidelity, which measures how faithfully important features contribute to a particular prediction.
We adopt the Descriptive Accuracy (DA) metric from~\cite{warnecke_evaluating_2020}, which measures for a given input sample $x$ how removing the $k$-most relevant features change the original neural network's prediction \mbox{$f_N(x) = y$ :} \mbox{$DA_k(x,f_N)=f_N(x|x_1=0,...,x_k=0)_y$}.

The idea behind this metric is that as relevant features are removed, accuracy should decrease as the classifier has less information to make the correct prediction.
The better an explanation, the faster the accuracy decreases as the removed features capture more context of the predictions.
Thus, explainability methods that show a more rapid decline in DA when removing key features provide better explanations than explainability methods with a more gradual decrease.

In context-less DGA classification, removing an input feature corresponds to removing a character from a domain.
Here, we consider two scenarios: (1) removing a character and thus reducing the total domain length, and (2) replacing a character with the padding symbol and thereby retaining the original domain length.
Both approaches have drawbacks: removing a character can have a greater impact on accuracy because it also affects the implicit feature of domain length.
On the other hand, preserving the domain length by replacing the character with the padding symbol may confuse a classifier, as the classifier was never faced with such samples during training.

Hence, we calculate the average DA for both scenarios and on all samples of DS\textsubscript{ex} for $k\in[1,10]$.
To derive a single score, we compute the Area Under the Curve (AUC).
The smaller the score, the better the explanations.

\paragraph{Results:}
In Table~\ref{tab:rlv_results}, we show the results for this criterion.
For further evaluation we choose \textit{integrated\textunderscore gradients} as it scores best when removing the top k-features and \textit{b-cos} as it achieves the best score in the second scenario.
In addition, we also select \textit{lrp.z\textunderscore plus} since it obtains the best scores when replacing features on the unmodified M-ResNet model.

\subsection{Sparsity}
An explanation is only meaningful if only a limited number of features are selected as the explanation result to make it understandable for a human analyst.
To measure the sparsity of an explanation, we follow the Mass Around Zero (MAZ) criterion proposed in~\cite{warnecke_evaluating_2020}.
First, for every sample, we calculate the relevance vector $r = (r_0,...,r_n)$, normalize the absolute entries of $r$ to the range $[0,1]$, and fit it to a half-normalized histogram $h$.
Then, we calculate the MAZ by \mbox{$MAZ(r) = \int_{0}^{1} h(x) dx$ for $r \in [0,1]$}.
Finally, we compute the AUC to derive a single score.
Sparse explanations have a steep increase in MAZ around zero and are flat around one because only few features are marked as relevant.
Conversely, explanations with many relevant features have a smaller slope close to zero.
Therefore, the higher the AUC score, the sparse the explanations.

\paragraph{Results:}
In the third column of Table~\ref{tab:rlv_results}, we show the results for this criterion.
We select \textit{lrp.alpha\textunderscore 2\textunderscore beta\textunderscore 1} for further evaluation as it shows the best sparsity for explanations.
However, high sparsity is only useful if the most relevant features are correctly determined.
Therefore, we also investigate $Sparsity * (1-Fidelity)$ and display the results in the fourth column.
Depending on the fidelity, \textit{integrated\textunderscore gradients} shows the most sparse explanations.

\subsection{Stability}
An explainability method is stable if it provides the same explanation for a given input over multiple runs.
Since we only evaluate white-box approaches which calculate the relevance vector deterministically, all methods are stable.

However, here we still want to evaluate the stability of the explainability methods over different model weights, i.e.,  whether the explainability methods calculate similar explanations via different model weights.
Assuming that all models converge to similar local optima, it is conceivable that they learn the same features that are similarly relevant to predictions of specific classes.
Note that this need not be the case as there may be multiple highly predictive features for a single class.
However, we believe this is an important criterion, as it is beneficial when deriving explanations in an operational environment that the security analyst is presented with similar explanations for the same classes after a model update, e.g., after the inclusion of a newly emerged malware family, as before the model update.
Otherwise, the new explanations would confuse rather than help the analyst.

The standard deviation of the f1-score across the four folds is low at 0.00552, which may indicate that the classifiers are converging to similar local optima.
To evaluate this criterion, we first compute the average of the standard deviation values (std) for each entry of a relevance vector across all folds for all domains.
Then, we average these values to derive a single score, with smaller values corresponding to more similar explanations across different model weights.

\paragraph{Results:} 
The fifth column of Table~\ref{tab:rlv_results} shows the results for this criterion.
The two methods which achieve the best results by far are \textit{deep\textunderscore taylor} and \textit{lrp.z\textunderscore plus}.

Both methods also achieve high fidelity scores (\textit{deep\textunderscore taylor} is second best in the feature remove setting and \textit{lrp.z\textunderscore plus} is best on the unmodified M-ResNet model in the feature replace setting), which may indicate that the models learn the same most predictive features for the same classes.

On the other hand, \textit{integrated\textunderscore gradients} achieves the best fidelity score in the feature remove setting and only performs moderately well in terms of stability.
This could be due to the fact, that in contrast to the other two methods, \textit{integrated\textunderscore gradients} shows a significantly higher sparsity, which could indicate that there may be multiple highly predictive feature combinations for the same classes.

We add \textit{deep\textunderscore taylor} to the list of methods to be evaluated further.
However, the results of this criterion should be treated with caution, as they depend heavily on what a model has learned.
Since we use the same models for all explainability methods, this criterion still allows us to compare explainability methods in terms of whether they provide similar explanations through different model weights.

\subsection{Efficiency}
We follow the definition of efficiency in~\cite{warnecke_evaluating_2020}, which states that a method is efficient if it does not delay the typical workflow of an expert. 
To evaluate this criterion, we measured and averaged the times to compute the explanations during the previous experiments.

\paragraph{Results:}
In the last column of Table~\ref{tab:rlv_results} we display the average time in seconds for computing a single explanation for a prediction.
All methods are sufficiently fast that we do not select any method based on this criterion.

\textit{B-cos}, \textit{integrated\textunderscore gradients}, and \textit{smoothgrad} are around on order of magnitude slower than the other approaches.
For \textit{B-cos} this is the case as the current implementation does not support batch calculations to derive explanations.
For \textit{integrated\textunderscore gradients} and \textit{smoothgrad} this is because we had to reduce the batch size of 2,000 samples to 200 due to higher RAM requirements of the algorithms.
Nevertheless, even without batch calculations all methods are sufficiently fast and would not delay the workflow of an expert.

\subsection{Comparison of Explainability Methods}
We briefly document our findings of using different explainability methods during our evaluations:

While \textit{lrp.alpha\textunderscore 2\textunderscore beta\textunderscore 1} often provides very sparse explanations, it occasionally seems to fail, sometimes just flagging features that argue against the prediction even though the classifier is very confident.

We cannot justify the loss of performance caused by the required adjustment to the state-of-the-art M-ResNet model for the explanations generated by \textit{b-cos}, since the explanations are not significantly different from the other methods.

The three best performing explainability methods through our study are \textit{deep\textunderscore taylor}, \textit{integrated\textunderscore gradients}, and \textit{lrp.z\textunderscore plus}.
All three can be used to explain the predictions of deep learning classifiers for the DGA classification use-case.
However, \textit{integrated\textunderscore gradients} seems to provide sparser explanations compared to the other two methods.

% !TEX root = ../paper.tex
\section{Additional ROC curves of the Real-World Study}
\label{sec:appendix}

\begin{figure}[!h]
	\centering
	\includegraphics[width=1.0\linewidth]{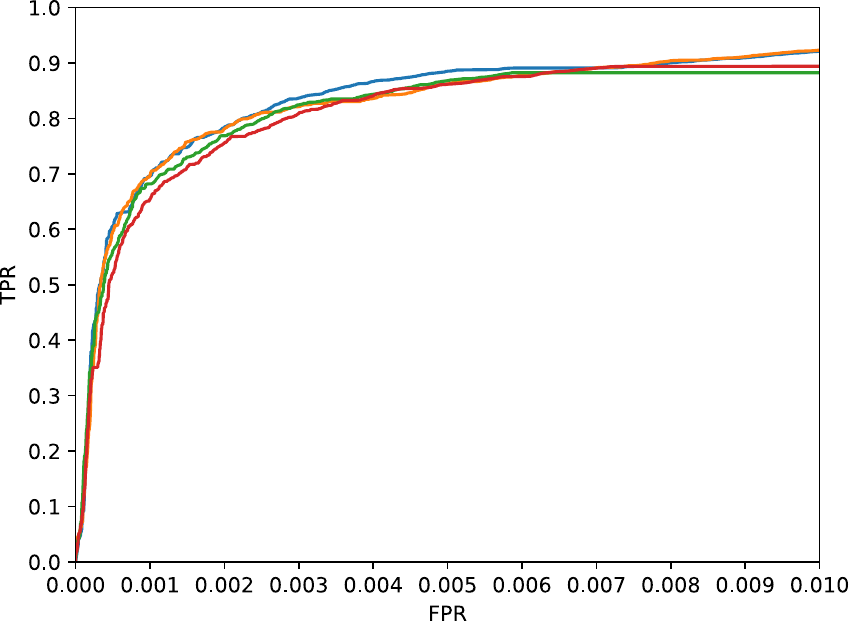}
	\caption{Experimental temporal bias afflicted ROC curves of the four bias-reduced e2LD classifiers of Section~\ref{sec:bias_mitigation_experiments}, evaluated against 311 million benign NXDs from the company network and DGA-domains generated by all 106 DGAs.}
	\label{fig:rw_roc}
\end{figure}

\end{document}